\documentclass [12pt]{article}
\usepackage[english]{babel}
\usepackage{graphicx,epsfig,amsmath,amsfonts,epstopdf}
\RequirePackage{graphicx}
\RequirePackage{epstopdf}

\def\*{\noindent}
\begin{document}
\title{One-dimensional model of cosmological perturbations: direct integration in the Fourier space}
\author{V.M.~Sliusar$^{1}$, V.I.~Zhdanov$^{2}$}
\date{}
\maketitle
\begin{center}
$^{1,2}$\textit{Astronomical Observatory, National Taras
Shevchenko University of Kyiv, Observatorna str.,3, Kiev 04053,
Ukraine }\\
$^1$ vitaliy.slyusar@gmail.com\\
$^2$ ValeryZhdanov@gmail.com
\end{center}
\begin{abstract}
We propose a method of calculation of the power spectrum of
cosmological perturbations by means of a direct numerical
integration of hydrodynamic equations in the Fourier space for a
random ensemble of initial conditions with subsequent averaging
procedure. This method can be an alternative to the cosmological
N-body simulations. We test realizability of this method in
 case of one-dimensional motion of gravitating matter pressureless shells.
In order to test the numerical simulations, we found an analytical
solution which describes one-dimensional collapse of plane shells.
The results are used to
study a nonlinear interaction of different Fourier modes. 
\end{abstract}
\section{Introduction}
Theoretical investigation of the cosmological inhomogeneity growth
presents one of the most serious challenges to computational
astrophysics. A number of problems arises on non-linear scales
when the density contrast cannot be assumed small. These include
simulations of the galaxy formation process, working out
predictions concerning the galactic environment (number of dwarf
satellite galaxies) and structure of their central regions (the
cusp-core problem)  either with cold dark matter (DM) or within
warm DM models
\cite{NFW1996,NFW1997,avila_2001,bode_2001,moore_2006,schneider_2012}.
Interesting possibility to obtain bounds on masses of DM particles
stems from observations of Ly-$\alpha$ forest (see, e.g.,
\cite{boyarsky}) and references therein), which requires accurate
calculations of the power spectrum of cosmological inhomogeneity
on kiloparsec scales.

Most developed computational techniques to study the cosmological
structure formation involve N-body simulations combined with the
smoothed particle hydrodynamics
\cite{springel_2005,brandbyge_2008,brandbyge_2008a}. Currently
performed simulations involve up to $10^9$ particles (see, e.g.,
\cite{brandbyge_2008,schneider_2012}).  These methods should be
tested in independent simulations.

On the other hand, some analytical and semi-analytical schemes
were proposed to study the matter power spectra on small scales
\cite{bernard_02,taruya_02,wong_08}.
 These
typically involve perturbative schemes after transition to the
Fourier-transformed hydrodynamical variables. There are techniques
dealing with correlation functions in the Fourier space
\cite{piet_08,lesg_piet_09}. These authors use some additional
{\textit apriori} suggestions in order to close the infinite chain
of correlation functions. The validity of these suggestions is not
evident. A comparison of different approaches can be found in
\cite{white_2009}.

In this paper we propose an alternative method, which uses a
direct integration of hydrodynamical equations  in the Fourier
space. In order to estimate workability of the method, as a first
step we consider a one-dimensional problem of hydrodynamical
evolution for a pressureless gravitating matter, i.e.
one-dimensional density shells. The integration is performed for
each realization from a random ensemble of initial data with
subsequent averaging procedure. In order to test the numerical
simulations, we found an analytical solution which describes
one-dimensional collapse of the plane shells (Section
\ref{analytical_solution}). This solution is used in case of
periodic initial data corresponding to a symmetric motion. In
section \ref{numerical_simulations} we write down the equations
for the Fourier coefficients and present some results for the
power spectrum obtained after  the statistical averaging.

\section{Implicit analytical solution}\label{analytical_solution}

In order to test numerical simulations, it is useful to have an
exact solution. In this section we obtain such a solution of
one-dimensional problem in the Lagrange variables. The
1-dimensional version of hydrodynamical equations (continuity,
Euler and Poisson equations) in case of a pressureless gravitating
fluid is:
\begin{equation}
\label{continuity} \frac{\partial \rho }{\partial t} +
\frac{\partial }{\partial x}\left( {\rho V} \right) = 0,
\end{equation}
\begin{equation}
\label{euler} \frac{\partial V}{\partial t} + V\frac{\partial
V}{\partial x} = - \frac{\partial \Phi }{\partial x},
\end{equation}
\begin{equation}
\label{eq6} \frac{\partial ^2\Phi }{\partial x^2} = 4\pi G\rho ,
\end{equation}
\noindent $V$ is the velocity, $\rho$ is the mass density, $\Phi$
is the gravitational potential, $G$ is the gravitational constant.
In order to pass to the Lagrangian description we introduce the
stream lines $X(\xi ,t)$:
\begin{equation}
\label{eq7}
\frac{dX}{dt} = V(X,t),\quad X(\xi ,0) \equiv \xi \quad .
\end{equation}
Equation (\ref{euler}) yields
\begin{equation}
\label{eq8}
\frac{d^2X}{dt^2} = E(X,t), \quad
E(x) = - \frac{\partial \Phi }{\partial x}
\end{equation}
Let for $t = 0$: $\rho (x,0) = \rho _0 (x), \quad V(x,0) = V_0
(x)$. Using (\ref{continuity}), (\ref{euler}) we see that $E(X(\xi
_1 ,t),t) - E(X(\xi _2 ,t),t) = const$
 is constant  along the stream lines. This is a consequence of
mass conservation in plane layer between $X(\xi _1 ,t)$ and $X(\xi _2,t)$. Then we obtain because of (\ref{eq8})
\begin{equation}
\label{eq9}
X(\xi _1 ,t) - X(\xi _2 ,t) = \xi _1 - \xi _2 + t\,\left[ {V_0 (\xi _1 ) -
V_0 (\xi _2 )} \right] - 2\pi G\,t^2\int\limits_{\xi _2 }^{\xi _1 } {dx}
\,\rho _0 (x).
\end{equation}
Formula (\ref{eq9}) allows to get a general solution of the
one-dimensional problem in Lagrangian coordinates. Further for
simplicity we deal with a symmetric motion (with respect to the
coordinate origin $\xi = 0$)  so that $X(0,t) \equiv 0$, $V_0 (0)
= 0$. In this case
\begin{equation}
\label{eq10}
X(\xi ,t) = \xi + t\,V_0 (\xi ) - 2\pi G\,t^2\int\limits_0^\xi {dx} \,\rho
_0 (x)
\end{equation}
The solution is valid during a limited time until
\begin{equation}
\label{eq11}
\frac{\partial X}{\partial \xi } = 1 + t\,{V}'_0 (\xi ) - 2\pi G\,t^2\rho _0
(\xi )\, \ne 0
\end{equation}
Transition to Euler variables is
\begin{equation}
\label{eq12} \rho (x,t) = \rho _0 (\xi )\left( {\frac{\partial
X}{\partial \xi }} \right)^{ - 1},\quad V(x,t) = V_0 (\xi ) - 4\pi
G\,t\int\limits_0^\xi {dx} \,\rho _0 (x)
\end{equation}
\noindent where $\xi = \xi (x,t)$ is defined implicitly as the solution of the equation:
\begin{equation}
\label{eq12implicit} x =X(\xi,t)
\end{equation}
In case of homogeneous initial conditions $\rho(x,0)=\rho _{h0}(x)
= const$, $V(x,0)\equiv V_{h0} (x) = H_0 x$, $ H_0$ is the "Hubble
constant", we get:
\[
x = \xi   \left( {1 + H_0 t - 2\pi G\rho _{h0} t^2} \right), \quad
\rho _h (x,t) = \frac{\rho_{h0} }{1 + H_0 t - 2\pi G\rho _{h0}
t^2},
\]
\begin{equation}
\label{homo} V_h (x,t) = x\frac{H_0 - 4\pi G\,t\,\rho _{h0} }{1 +
H_0 t - 2\pi G\rho _{h0} t^2}
\end{equation}
\noindent or
\begin{equation}
\label{eq14}
\rho _h (x,t) = \frac{\rho _{h0} }{R(t)},\quad V_h (x,t) = x\frac{\dot
{R}(t)}{R(t)},
\quad
R(t) = 1 + H_0 t - 2\pi G\rho _{h0} t^2.
\end{equation}

To consider deviations from the homogeneous background, we impose
periodic boundary conditions such that the initial conditions ($t
= 0$) are as follows:
\[
V_{0} (\xi )=H_{0} \left[ {\xi +\sum\limits_{\substack{n=1
}}^\infty {\frac{a_{n}^0 }{k_{n} }} \sin (k_{n} \xi )}\right],
\,\, \rho_{0} (\xi )=\rho_{h0} \left[ {1
+\sum\limits_{\substack{n=1 }}^\infty {b_{n}^{\,0}} \cos (k_{n}
\xi )} \right],
\]
where
%$ a_n = (a_{ - n} )^{\ast} ,\quad b_n = (b_{ - n} )^{\ast}$,
 $L$ is  the periodic "box" size, $k_n = 2\pi n / L$.

Substitution of the initial conditions into
(\ref{eq10}-\ref{eq12}) yields

\[
X(\xi,t)=R(t)\xi+tH_0
\sum\limits_{n=1}^{\infty}\frac{a_n^0}{k_n}\sin(k_n\xi)-\frac{3}{4}\left(t\,H_0\right)^2
\sum\limits_{n=1}^{\infty}\frac{b_n^{\,0}}{k_n}\sin(k_n\xi)\,,
\]
\[ \rho(x,t)=\frac{\rho_{h0} \left[ {1 +\sum\limits_{\substack{n=1
}}^\infty {b_n^{\,0}} \cos (k_{n} \xi )} \right]}{R(t)+tH_0
\sum\limits_{n=1}^{\infty} a_n^0
\cos(k_n\xi)-\frac{3}{4}\left(t\,H_0\right)^2
\sum\limits_{n=1}^{\infty} b_n^{\,0} \cos(k_n\xi)},
\]

The solution shows a singular growth (collapse of one-dimensional
gravitating layers) as the condition (\ref{eq11}) is violated.

At the end of this Section we note that in spite of the exact form
of the solution given by (\ref{eq10}-\ref{eq12}) we cannot avoid a
numerical work when we pass to the Euler variables and then to the
Fourier representation. So we refer to the method of this Section
as "semi-analytical".

\section{Numerical simulations}\label{numerical_simulations}
Furthermore for the homogenous background we denote $\mathcal
H(\tau ) \equiv dR/dt$,  $\tau $ is the "conformal time":
$dt=R(\tau)d\tau$, and $y=x/R(\tau)$ is the comoving spatial
coordinate. After some calculation on account of (\ref{eq14}) we
get
\[
\tau=\frac{1}{2H_0}ln\left[\frac{1+3H_0 t}{3(2-H_0 t)}
\right],\,\, \mathcal H(\tau )=\frac{dR}{dt}=
\frac{1}{R}\frac{dR}{d\tau } = - 2H_0 \tanh (H_0 \tau ).
\]

Hereafter $\delta$ is the density contrast, $\theta = \partial v /
\partial y$, $v$ is the peculiar velocity.
 Taking into account Poisson equation (\ref{eq6}) we get:
\begin{equation}
\label{eq15} \frac{\partial ^2\phi }{\partial y^2} = \alpha R(\tau
)\delta ,\quad \alpha = 4\pi G\rho _{h0} ,
\end{equation}
In terms of conformal $\tau $ and comoving $y$ the hydrodynamic
equations can be written as:
\begin{equation}
\label{eq16} \frac{\partial \delta }{\partial \tau } + \theta +
\frac{\partial }{\partial y}\left( {v\delta } \right) = 0
\end{equation}
\begin{equation}
\label{eq17} \frac{\partial \theta }{\partial \tau } + H{\kern
1pt} \theta + \frac{\partial }{\partial y}\left( {\theta v}
\right) = - \alpha R\delta ,
\end{equation}

We proceed to deal with the Fourier coefficients in the symmetric
one-dimensional case.
\begin{equation}
\label{eq18}
\delta (x,\tau ) = \sum\limits_{n = - \infty }^\infty {b_n (\tau )} \exp
(ik_n x),
\quad
k_n = 2\pi n / L,
\end{equation}
\begin{equation}
\label{eq19}
\theta (x,\tau )=\sum\limits_{n=-\infty }^\infty {a_{n} (\tau )} \exp
(ik_{n} x),\quad v(x,\tau )=\sum\limits_{n=-\infty }^\infty {\frac{a_{n}
(\tau )}{ik_{n} }} \exp (ik_{n} x).
\end{equation}
The reverse transformation is:
\begin{equation}
\label{eq20}
b_{n} (\tau )=L^{-1}\;\int\limits_0^L {dx} \,e^{ik_{n} x}\delta (x,\tau ),
\quad
a_{n} (\tau )=L^{-1}\;\int\limits_0^L {dx} \,e^{ik_{n} x}\theta (x,\tau )
\quad
\end{equation}
We assume $a_0 = 0,\quad b_0 = 0$ at $t=0$ then it is easy to see
from (\ref{eq16},\ref{eq17}) that these equalities are fulfilled
for all $t>0$.

The equations for the Fourier coefficients take on the form:
\begin{equation}
\label{eq21}
\frac{da_{n} }{d\tau }+H(\tau )a_{n} +\alpha R(\tau )b_{n}
+n\sum\limits_{\substack{p=-\infty \\ p\ne 0} }^\infty {\frac{a_{p} a_{n-p} }{p}} =0,\quad n=\pm 1,\pm 2,...
\end{equation}
\begin{equation}
\label{eq22}
\frac{db_{n} }{d\tau }+a_{n} +n\sum\limits_{\substack{p=-\infty \\ p\ne 0}}^\infty {\frac{a_{p} b_{n-p} }{p}} =0,\quad n=\pm 1,\pm 2,...
\end{equation}
The numerical solution of the equations (\ref{eq21}), (\ref{eq22})
was performed using the 4-th order Runge-Kutta method after a
truncation of the infinite chain of coefficients $a_n,b_n$.
Calculations were carried out by the specially written GPGPU code
using OpenCL SDK by AMD. The time, required to calculate $a_{n}$
and $b_{n}$, in case of 256 values of $n$ (points over $k$) for
single realization of initial conditions, is 10 seconds. This is
considerably faster than direct using of the semi-analytical
solution of Section \ref{analytical_solution}. It is important to
note that it is easy to extend the corresponding algorithms to the
3-D case.

\begin{figure}[H]
\includegraphics[width=140mm]{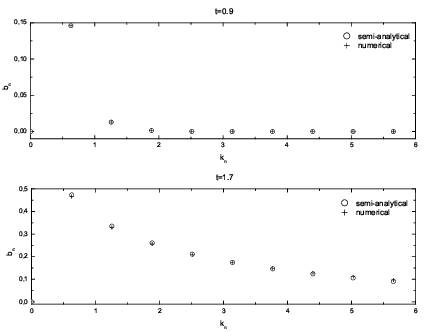}
\vskip-3mm\caption{Coefficients $b_{n}(t)$ for $t=0.9$ (top) and
$t=1.7$ (bottom)  determined by semi-analytical and numerical
methods with the only nonzero initial values $b_{\pm 1}(0)= 0.1$.}
\label{fig1}
\end{figure}

We calculated coefficients $b_{n}$ of the density contrast as a
function of $t$ by means of analytical and numerical methods with
the same initial conditions. On the Fig. \ref{fig1} these
coefficients are presented for $t = 0.9$ and $t = 1.7$ for both
methods. For larger $t$ we observe an infinite growth (for finite
time) that corresponds to collapse of some plane gravitating
shells due violation of condition (\ref{eq11}). Correspondingly,
the difference between two methods, that reflects the error of
calculation, increases for greater $t$ and greater $n$. For
example, in order to look how the perturbation propagates from
small wavenumbers to larger ones, we considered the following
initial conditions: $b_{n}(0) = 0$ where integer $n$ varies $
-128$ to $128$ except $n=\pm 1$; $b_{\pm 1}(0)=b_{\pm1}^{\,0}/2 =
0.1$; all $a_n(0)=0$. For $t = 0.2$ or $t = 0.5$ the difference
between the values calculated by different methods is less than
1{\%}, and for $t = 1.7$ the difference changes from 1.5{\%} to
6.7{\%} as $k_n = 2\pi n / L$ increases from 0.6 to 6. Larger
$k$-interval is presented on Fig. \ref{fig2}. The next figure
presents the power spectrum obtained by averaging of the solutions
for the ensemble of initial data with uniform distribution of
$b_{\pm 1} (0)$,  $<b_{\pm 1}^{\,2}(0)>=0.5$ . We observe the
growth of dispersion, which is explained as follows: as $t$ grows,
some of realizations of the ensemble of solutions (with larger
$|b_{\pm 1} (0)|$) enter the region which is close to the
singularity.

\begin{figure}[H]
\includegraphics[width=140mm]{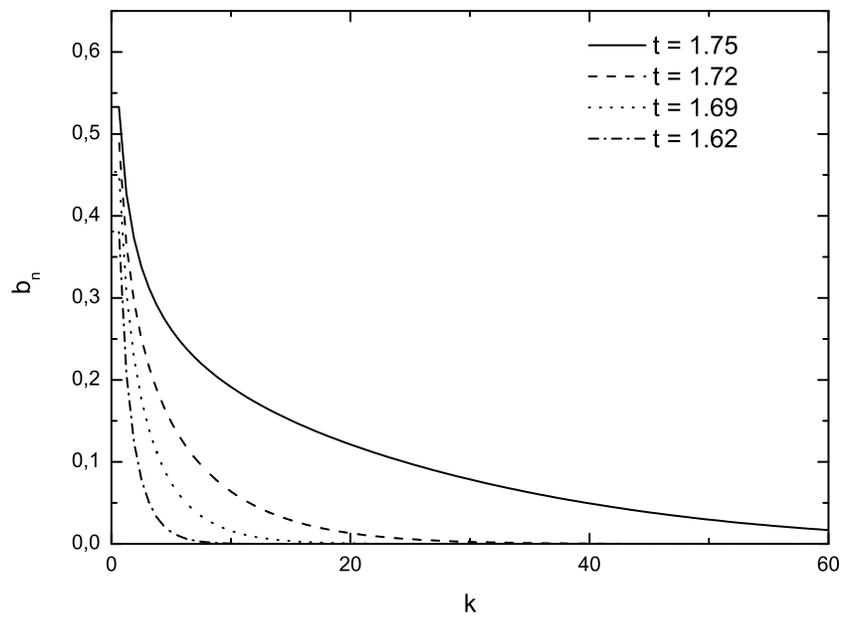}
\vskip-3mm\caption{The evolution of perturbations over $k$ for the
initial conditions as described on Fig.~\ref{fig1}. } \label{fig2}
\end{figure}
\begin{figure}[H]
\includegraphics[width=140mm]{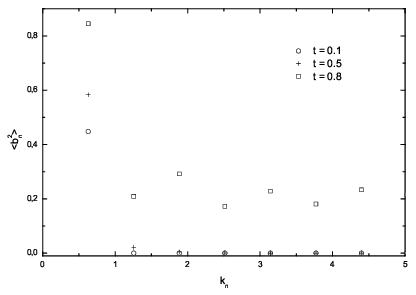}
\vskip-3mm\caption{Power spectrum $<b_{n}^{\,2}>$ calculated for
100 realizations of randomly generated independent initial
conditions. } \label{fig3}
\end{figure}
\section{Conclusions}
We present a new approach to investigation of the cosmological
inhomogeneity by means of the direct integration of hydrodynamic
equations in the Fourier space. At the moment we studied a
one-dimensional hydrodynamical evolution of cold (pressureless)
gravitating matter. The numerical integration has been fulfilled
for a random ensemble of initial conditions with subsequent
averaging procedure to get the power spectrum of the density
contrast. We used the GPGPU instead of the classic CPU because the
problem can be easily processed in parallel environment.

The numerical simulations have been tested using the analytic
solution that describes the one-dimensional collapse of
gravitating shells. The density contrast shows a propagation of
perturbations from smaller wavenumbers to larger ones. The
evolution in time ends with a singular growth of the density
contrast. Correspondingly, we point out a significant growth of
dispersion of the power spectrum in the non-linear region.

We consider our results as a first step to the simulations of
cosmological inhomogeneity growth in the cold matter that could be
an alternative to the cosmological N-body simulations. We expect
that our approach will be especially effective in the weakly
nonlinear regime. The next step will be an implementation of the
three-dimensional case of the problem, which is technically
similar to the one dimensional problem. The trial runs of our
method allow us to think that it could be really used for power
spectrum calculations in the 3-D case with realistic requirements
to the computer time.

{\textit {Acknowledgements}}. This work has been supported in part
by Swiss National Science Foundation (SCOPES grant 128040).


\begin{thebibliography}{99}%

\bibitem{NFW1996}
J.F.~Navarro, C.S.~Frenk, S.D.M.~White, \emph{The structure of
cold dark matter halos}. ApJ, \textbf{462},  P.563-575 (1996).


\bibitem{NFW1997}
J.F.~Navarro, C.S.~Frenk, S.D.M.~White, \emph{A universal density
profile from hierarchical clustering}.  ApJ,  \textbf{490},
P.493-508 (1997).

\bibitem{avila_2001} V. Avila-Reese, P. Col\'iin, O. Valenzuela, E. D'Onghia,
C. Firmani, \emph{Formation and Structure of Halos in a Warm Dark
Matter Cosmology}. ApJ, \textbf{559}, Is.2, P.516-530 (2001).

\bibitem{bode_2001} P.~Bode, J.P.~Ostriker, N.~Turok, \emph{Halo Formation in Warm Dark Matter
Models}.  ApJ, \textbf{556}, Is.1, P.93-107 (2001).

%\bibitem{gilmore_2007} G. Gilmore, M.I.Wilkinson, R.F.G.Wyse, et al., \emph{The Observed Properties of Dark Matter
%on Small Spatial Scales}. ApJ. {\bf 663}, 948 (2007).


\bibitem{moore_2006} T.~Goerdt, B.~Moore, J.I.~Read, J.~Stadel,
M.~Zemp, \emph{Does the Fornax dwarf spheroidal have a central
cusp or core?} MNRAS,   \textbf{ 368}, Is.3, P.1073-1077 (2006).


%\bibitem{lovell_2011}M.~Lovell, V.~Eke, C.~Frenk, et al. \emph{The Haloes of Bright
%Satellite Galaxies in a Warm Dark Matter Universe}, ArXiv e-prints
%(Apr., 2011) [1104.2929].

\bibitem{schneider_2012} A.~Schneider, R.E.~Smith, A.V.~Macci\`o, B.~Moore,
\emph{Non-linear evolution of cosmological structures in warm dark
matter models}. MNRAS, \textbf{424}, Is.1, pp. 684-698 (2012).

\bibitem{boyarsky} A.~Boyarsky, J.~Lesgourgues, O.~Ruchayskiy, M.~Viel,
 \emph{Lyman-$\alpha$ constraints on warm and on warm-plus-cold dark matter
 models}.  JCAP, \textbf{05} id. 012 (2009).

\bibitem{springel_2005} V.~Springel, \emph{The cosmological simulation code GADGET-2}.
 MNRAS \textbf{364}, P.1105-1134 (2005).

\bibitem{brandbyge_2008} J.~Brandbyge, S.~Hannestad, T.~Hangbolle, B.~Thomsen, \emph{The
Effect of Thermal Neutrino Motion on the Non-linear Cosmological
Matter Power Spectrum}. JCAP, \textbf{08}, id. 020, 16 pp. (2008).



\bibitem{brandbyge_2008a} J. Brandbyge and S. Hannestad, \emph{Grid Based Linear Neutrino
Perturbations in Cosmological N-body Simulations}.  JCAP,
\textbf{05}, id.002 (2009).


\bibitem{bernard_02} F. Bernardeau, S. Colombi, E. Gazta\~naga, R. Scoccimarro.
\emph{Large-scale structure of the Universe and cosmological
perturbation theory}. Phys. Rep. \textbf{ 367}, Is.1-3, P.1-248
(2002)


\bibitem{taruya_02} A. Taruya and T. Hiramatsu. \emph{A Closure Theory for Nonlinear
Evolution of Cosmological Power Spectra}. Astrophys. J.
\textbf{674}, Issue 2, pp. 617-635 (2008).

\bibitem{wong_08} Y. Y. Y. Wong, \emph{Higher order corrections to the large scale
matter power spectrum in the presence of massive neutrinos}. JCAP,
\textbf{10}, id. 035, 24 pp. (2008).


\bibitem{piet_08} M.~Pietroni. \emph{Flowing with time: a new approach to non-linear cosmological perturbations}.
 JCAP, \textbf{10}, id.19 (2008), 19 pp.

\bibitem{lesg_piet_09}J.Lesgourgues, S.Matarrese, M.Pietroni, A.Riotto. \emph{Non-linear power
spectrum including massive neutrinos: the time-RG flow approach}.
JCAP, \textbf{06}, id.017 (2009).

\bibitem{white_2009} J.~Carlson, M.~White, N.~Padmanabhan. \emph{A critical look
at cosmological perturbation theory techniques}. Phys.Rev.
\textbf{D80}, 043531 (2009)


\end{thebibliography}
\end{document}